# Real-time Terahertz Wave Channeling via Multifunctional Metagratings: A Sparse Array of All-Graphene Scatterers

SAHAR BEHROOZINIA, HAMID RAJABALIPANAH, ALI ABDOLALI

[1]Applied Electromagnetic Laboratory, School of Electrical Engineering, Iran University of Science and Technology, Tehran, Iran.

**Acquiring full control over a large number of diffraction orders can be strongly attractive in the case of realizing multifunctional devices such as multichannel reflectors. Recently, the concept of metagrating has been introduced which enables obtaining the desired diffraction pattern through a sparse periodic array of engineered scatterers. In this letter, for the first time, a tunable all-graphene multichannel meta-reflector is proposed for operating at terahertz (THz) frequencies. In the super cell level, the designed metagrating is composed of three graphene ribbons of different controllable chemical potentials which can be regarded as a five-channel THz meta-reflector. Several illustrative examples have been presented in which by choosing proper distribution of DC voltages feeding the ribbons, our design can realize different intriguing functionalities such as anomalous reflection, retro-reflection, and three-channel power splitting within a single shared aperture and with high efficiency. This work paves the way toward designing highly-efficient and tunable THz multichannel meta-reflectors with many potential applications in photonics and optoelectronics.**

Devising optical systems capable of full control over the distribution of electromagnetic (EM) waves has been a growing research hotspot in the past decades. Recently, with the advent of metasurfaces, 2D reductions of volume metamaterials, it was shown that by imparting a suitably tailored transverse phase change over an ultrathin surface, diverse field transformations can be achieved obeying the generalized Snell's law [1]. A vast variety of designs based on this particular class of metasurfaces called as gradient metasurfaces have been reported in [2-5]. More recently, by revisiting the problem of full power coupling from an incident plane wave with a given angle to a reflected wave propagating towards an arbitrary direction, it was proved that the gradient architectures suffer from inevitable efficiency limitations, especially at extreme angles [6]. On the way toward perfect wave manipulating schemes, inspired by traditional diffraction grating principles, the concept of metagrating was revolutionary introduced in [7]. It has been demonstrated that one can acquire unprecedented wave transformation efficiencies through the use of a sparse periodic array (non-gradient) of suitably engineered scatterers (only a few subwavelength meta-atoms in each period). Meanwhile, they substantially overcome the significant implementation challenges such as high resolution fabrication demands associated with the metasurfaces imposed by aggressive discretization requirements.

In [7-9], by considering the single meta-atom per supercell and the substrate thickness as the degrees of freedom, a maximally discretized metagrating was found to have the ability of eliminating specular reflection and gaining perfect beam splitting and anomalous reflection. In [10], the authors experimentally demonstrated that using an aggressively-discretized Huygens' metasurface with only two simple elements per period, required for suppressing the specular reflection, the perfect anomalous reflection phenomenon can be guaranteed over a wide angular and frequency range. Realizing the perfect abnormal reflection along the 1$^{st}$ diffraction harmonic and suppressing both the zeroth and minus first diffraction orders by utilizing two line currents per supercell, was also outlined in [11]. *Popov et al.*, have illustrated that a metagrating occupied with a number of meta-atoms per supercell, equal to the number of plane waves scattered in the far field domain, can be elaborately exploited for flexible control of the diffraction pattern [12]. Besides, a straightforward and robust theory for designing the involved meta-atoms was reported in the same reference. Although the abovementioned proposals investigated engineered surfaces that can efficiently manipulate the EM wavefronts, the yield meta-devices expose only single functionality. Asadchy *et al.* introduced a concept of multichannel functional metasurfaces, which are able to control incoming and outgoing waves in a number of propagation directions [13]. Nonetheless, the multichannel response of the proposed flat reflector was obtained at the expense of using complex design and fabrication procedures involving ten rectangular conducting patches over a metallic plane. The possibility of designing multifunctional metagratings has been also inspected in [14], wherein multiple functionalities have been enabled through a single interface consisting of two subwavelength graphene/metal hybrid scatterers, whose characteristics can be electrostatically tuned by external biasing voltages. However, the presented study suffers from a sophisticated architecture, low diversity of the associated functionalities, and limited number of the prepared channels. and Moreover, based on the combination of the array antenna theorem the metagrating concept, three different terahertz (THz) multichannel reflectors with no more than three particles per period were designed and numerically demonstrated in [15]. Nevertheless, the functionality of the presented designs remains unchangeable once the structure is fabricated and no reconfigurability has been addressed. Eventually, to the best of the authors' knowledge, there has been no report on the metagrating designs which have the ability to dynamically tailor the power distribution of more than three channels, yet.

Here, for the first time, we propose a multichannel reconfigurable metagrating comprising of all-graphene scatterers to realize diverse scattering functionalities with merely modulating the biasing voltages in a shared THz aperture. Each supercell consists of three equi-width graphene ribbons with different chemical potentials. The power distribution over all scattering channels can be instantaneously altered by varying the DC voltages applied to each graphene ribbon. Based on what was outlined in [15], this number of meta-atoms per period can provide sufficient degrees of freedom to attain full control over the available diffraction channels at will. Compared with the previous multifunctional designs [14-15], the designed architecture is regarded as the first all-graphene proposal with a simpler geometry while exposing a higher number of controllable channels for THz wave manipulation. We numerically demonstrate that, our design can successfully act as an anomalous reflector, retroreflector, and three-channel power splitter when excited from specific ports.

Fig. 1(a) depicts the schematic of the proposed tunable metagrating in the supercell level, consisting of three graphene ribbons placed on top of a grounded dielectric layer. All the graphene ribbons have the same widths and their optical properties are solely controlled by using external electrostatic biasing voltages. A supporting dielectric layer ($\varepsilon_r = 2.1$) with the thickness of $h$ is utilized as the core substrate which is terminated by a perfect electric conductor (PEC) to avoid energy transmission across the frequency band of study. TM-polarized illuminations are assumed throughout this letter. Since the periodicity of the graphene-based supercell along the y direction is deliberately chosen to be larger than the operating wavelength, the designed metasurface can be optically regarded as a 1D multichannel reflector (Figs. 1(b), (c)). The corresponding y-directed diffraction channels are oriented toward the angles determined through the Floquet-Bloch analysis:

$$\theta_n = \arcsin(\sin\theta_i + n\lambda/P_y) \qquad (1)$$

in which, $n = 0, \pm 1, \pm 2$ refers to the harmonic (channel) number. Upon illuminating by normal incidences ($\theta_i = 0°$), if $\lambda < P_y < 2\lambda$, only three propagating channels are opened with the orientation angles of $\theta_{r0} = 0°$ and $\theta_{r\pm 1} = \pm\arcsin(\lambda_0/P_y)$, (the diffraction angles are defined in the clockwise direction with respect to the z-axis) wherein the subscript indices correspond to the harmonic number $n$. Such a configuration is displayed in Fig. 1(b). The same ports are also excited when the metagrating is shinned by oblique incidences at $\theta_i = \pm\arcsin(\lambda_0/P_y)$. Nevertheless, being illuminated from $\theta_i = \pm\arcsin(0.5\lambda_0/P_y)$ directions as shown in Fig. 1(c), the metagrating exposes only two open ports along $\theta_{r-1,0} = \pm\arcsin(0.5\lambda_0/P_y)$. Thus, although the reflector has five channels, three of its channels (Ports 1, 3, and 5) are isolated from the other two (Ports 2 and 4). We choose the y-directed periodicity of the metagrating as $P_y = 1.412\lambda_0$ so that the five propagating channels will be oriented along $0°$ (Port 3), $\pm 20.7°$ (Ports 2 and 4), and $\pm 45°$ (Ports 1 and 5). In what follows, we present different functionalities of the designed meta-grating by focusing over the scattering parameters of the open diffraction channels when illuminated from $\theta_i = 0°$ and $\pm 45°$ directions. Then, we show that by using proper electrostatic voltages feeding the graphene ribbons, the same meta-grating can simultaneously serve as a multi-angle reconfigurable retroreflector when excited from the two other channels, i.e. $\theta_i = \pm 20.7°$.

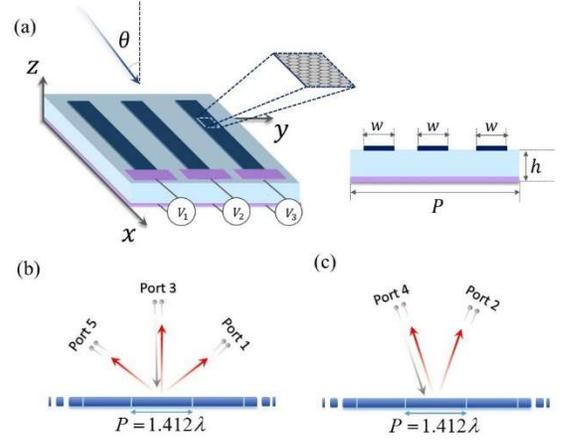

Fig. 1. Schematics of (a) the designed supercell consisting of three graphene ribbons fed by different DC voltages and our five-channel metagrating as a combination of isolated (b) three-channel and (a) two-channel wave manipulating networks, respectively.

As the first illustrative example, the metagrating is elaborately designed to perfectly deflect an incident wave coming from Port 3 toward the Port 5 with the tilt angle of $\theta_r = \theta_{-1} = -45°$ [see Fig. 2 (a)]. As mentioned above, shinning the proposed meta-grating from each of Ports 1, 3 and 5 yields three open floquet channels. Therefore, N=3 and we need three degrees of freedom to manipulate the amplitude of these modes and achieve the desired diffraction pattern at will. Our proposed meta-grating with three individual graphene ribbons can provide enough flexibility allowing us to get any specific design goal. Here, the graphene ribbons have been considered as a 2D boundary with a complex surface conductivity $\sigma_s$ obeying the Kubo formula [16], [17]:

$$\sigma_s(\omega, E_f, \Gamma, T) = \frac{e^2 k_B T}{j\pi\hbar^2(\omega - 2j\Gamma)}\left[\frac{\mu_c}{K_B T} + 2\ln(e^{-\mu_c/k_B T} + 1)\right] \quad (2)$$
$$+ 2\ln(e^{-\mu_c/k_B T} + 1)] + \frac{e^2}{4j\pi\hbar}\ln\left[\frac{2|\mu_c| - (\omega - 2j\Gamma)\hbar}{2|\mu_c| + (\omega - 2j\Gamma)\hbar}\right]$$

Herein, $e$ is the electron charge, $E_f$ indicates the fermi energy, $\mu_c$ denotes the chemical potential, $\hbar$ remarks the Plank constant, $\omega$ is the angular frequency, $T = 300K$ is the environment temperature and, $\tau$ represents the relaxation time (1 ps throughout this letter). The surface conductivity of each graphene ribbon can be individually controlled by using external electrostatic fields separately applied between the ground plane and the conductive pads touching the graphene ribbons.

A full-wave commercial software, CST Microwave Studio, has been established to numerically model the designed metagrating in which the periodic boundary condition and the Floquet ports have been utilized in our simulations. A comprehensive parametric study has been accomplished to find the best structural parameters including the suitable chemical potentials ($\mu_{c1}, \mu_{c2}, and\ \mu_{c3}$) and the other design parameters i.e. the widths ($w$) of the graphene ribbons and the thickness of the dielectric ($h$). The chemical potential of each graphene ribbon can be initially set to a certain value via strong doping of the graphene sheets. Dynamic changes in the chemical potentials can then be accessible through feeding the ribbons by the external electrostatic fields [16, 18]. Fig. 2(a) displays the results for the case in which the designed metagrating acts as a highly-efficient anomalous reflector (Tx: Port 3 and Rx: Port 5) operating at the frequency of $f=5\ THz$. The optimum values of $w, h, \mu_{c1}, \mu_{c2}$, and $\mu_{c3}$ parameters for this case are 8.5 $\mu m$, 10 $\mu m$, 0.02 $eV$, 1.02 $eV$, and 1.49 $eV$, respectively. As can be observed, the normally incident power coming from the Port 3 has been channeled into the Ports 1, 3, and 5 with the proportion of 0.5%:5%:77%, respectively. It should be noted that about 17% of the incident power has been dissipated due to the ohmic loss occurring in the graphene material. Indeed, only 5.5 % of the incident power is wasted because of the parasitic scattering. To the best of our knowledge, this anomalous reflection efficiency is still higher than that of the previously reported reconfigurable gradient metasurfaces [19]. As can be deduced from Fig. 2(b), if the designed metagrating is excited from θ$_i$=−45° (Port 5), it will redirect the incident fields into the boresight direction (Port 3) with a high efficiency. At the same time (see Fig. 2(c)), the metagrating will play the role of a retroreflector when illuminated from Port 1 as if the incident beam faces an oblique optical mirror. The both deductions can be interpreted by the energy conversation and reciprocity principles. The numerical results corresponding to the scenarios in which Ports 5 and 1 launch the input signals are demonstrated in Figs. 2(a)-(c), respectively. As the inset of these figures show, the calculated efficiencies are as high as 1.7%:77%:5% and 80%:0.5%:1.7%, respectively. The proposed metagrating can also effectively operate in a different scenario, if we change the biasing voltages applied to the first and last graphene ribbons. It will couple the power carried by a normally incident wave coming from Port 3 into Port 5, as illustrated in Fig. 3(a). Keeping the reciprocity theorem in mind, if we excite the same aperture from Port 1, the metagrating will re-direct the whole reflected power, excluding losses, into the normal direction (Port 3) (Fig. 3(b)). Indeed, Port 5 will be isolated meaning that the total energy coming from this port is forced to be reflected back, as shown in Fig. 3(c). The corresponding results can be found in Figs. 3(a)-(c), respectively, from which the wave transformation efficiencies are noticed as high as 77%:5%:0.5%, 5%:77%:1.7%, and 1.7%:0.5%:80%, respectively.

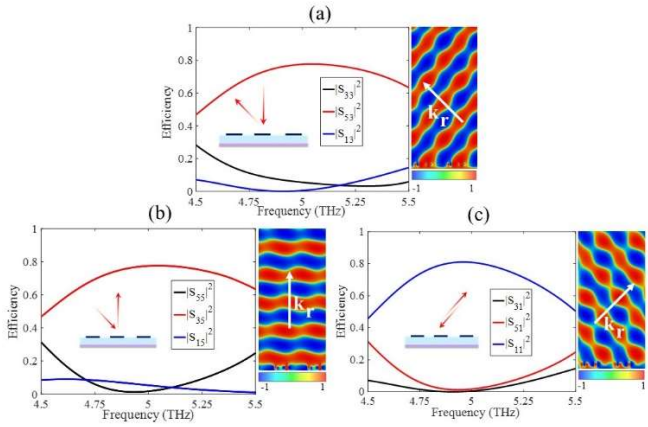

Fig. 2. The scattering parameters of the metagrating operating in different scenarios of (a) anomalous reflection from Port 3 to Port 5, (b) anomalous reflection from Port 5 to Port 3 and (c) retroreflection at Port 1. The corresponding scattered fields are plotted in the inset of figures.

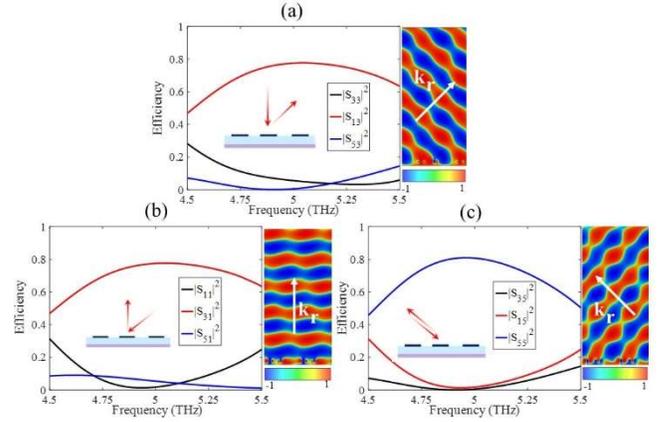

Fig. 3. The scattering parameters of the metagrating operating in different scenarios of (a) anomalous reflection from Port 3 to Port 1, (b) anomalous reflection from Port 1 to Port 3 and (c) retroreflection at Port 5. The corresponding scattered fields are plotted in the inset of figures.

To further illustrate the flexibility of our proposed reconfigurable architecture, we also design a three-channel THz power splitter through the mere modulation of the chemical potentials. In the first case, we target a metagrating equally splitting the input signal of Port 3 between Ports 1 and 5. The three chemical potentials obtained from our optimization procedures are 1.28 $eV$, 0 $eV$ and 0 $eV$, respectively. The numerical results are shown in Fig. 4(a) indicating that when our design is excited from Port 3, the power channeled to each channel is 40%, 6.7%, and 40% (pertaining to the Ports 1, 3, and 5, respectively). Obviously, regarding the lossy characteristic of the graphene material, about 13.3% of the incident power is trapped and absorbed in the graphene ribbons. Therefore, only 6.7% of the input power will be involved in parasitic scattering. It may well be noted that, if the same metagrating be excited from the two other channels, it cannot act as a power splitter and the scattered energy will be distributed in all three Ports. For instance, when an incident wave impinges on the metagrating from Port 1, the power ratios coupled into the Ports 1, 3, and 5 are 40% ,25% , and 25%, respectively. This observation can be also deduced from the basic theory of the multiport network, stating that the lossless, passive, three–port splitters which are matched at all channels cannot exist [20]. Here, we overcome this shortcoming by exploiting the tunability of the graphene-based meta-atoms. In what follow, we demonstrate that by suitably biasing the graphene ribbons, the proposed metagrating can act as a power splitter in all available three ports. We re-organize the chemical potentials of the graphene ribbons so that the THz waves coming from Port 5 is divided equally between Ports 1 and 3 where no power is almost reflected back into Port 5. The optimum chemical potentials for this purpose are obtained as 0.2$eV$, 0.68$eV$, and 1.05 $eV$, respectively. The numerical results plotted in Fig. 4(b) remark that when the metagrating is illuminated from Port 5, the incoming energy is splitted towards 45° (Port 1), 0° (Port 3), and -45° (Port 5) with the proportion of 41%:42%:4%. The rest power is absorbed by the graphene material. In order to have a power splitter excited from Port 1, one just need to alter the first and third biasing voltages obtained in the previous case. The numerical results for this case are depicted in Fig. 4(c) where in accordance with our expectations, they indicate that when the metagrating is launched from Port 1, the energy is splitted towards -45(Port 5) ,0 (Port 3) and 45 (Port 1) with the proportion of 4%:41%:42%. The rest power is absorbed by the graphene material.

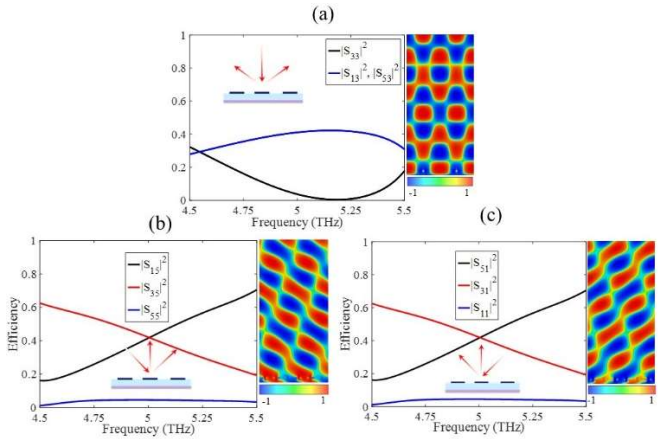

Fig. 4. The scattering parameters of the metagrating operating in the three-channel splitting mode excited from (a) Port 3, (b) Port 5, and (c) Port 1. The corresponding scattered fields are plotted in the inset of figures.

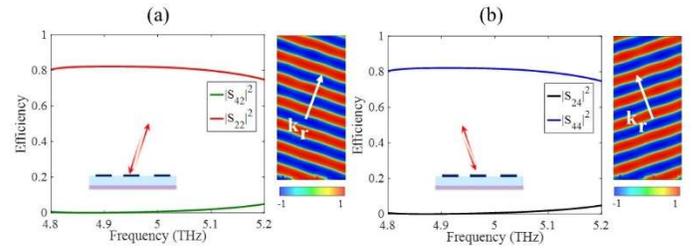

Fig. 5. The scattering parameters of the multichannel metagrating playing the role of a reconfigurable retroreflector when seen from both (a) Port 2 and (b) Port 4. The corresponding scattered fields are plotted in the inset of figures.

As the final demonstration, we intend to picture the capability of our multichannel metagrating in control of the rest two ports (i.e. Port2 (20.7°) and Port4 (-20.7°)), which are isolated from the other three ports studied in the previous examples. In this case in which only two propagating floquet modes i.e., 0 and -1 exist, we design the metagrating so as to perfectly reflect the incident energy coming from $\theta_i = 20.7°$ (Port 2) into the same direction. Consequently, we need two degrees of freedom to tailor the amplitudes of these two modes and our structure with three chemical potentials can provide enough flexibility to eliminate the unwanted diffraction order 0 and set the reflection amplitude of the desired diffraction mode -1 to 1. The optimized chemical potentials for this purpose can be listed as 1.32 $eV$, 0.6 $eV$, 1.32 $eV$, respectively. Fig. 5(a) shows the results corresponding to the retro-reflection functionality of Port 2 (20.7°), disclosing that 81% of the incident power will be reflected back into the input port and 0.7% is coupled to Port 4. About 18.3% of the incident power is also absorbed due to the ohmic losses in the graphene ribbons. Because of the reciprocity and the negligible dissipated power, Port 4 is also isolated and when we excite the structure from Port 4, the metagrating behaves as a retroreflector, at the same time. As can be deduced from Fig. 5(b), the result for this case is the same as the case of exciting Port 2.

In conclusion, a reconfigurable multichannel metagrating made of all-graphene scatterers was proposed in this letter for the first time to realize diverse THz scattering functionalities in a single shared aperture. The designed metagrating was composed of three equi-width graphene ribbons with different chemical potentials in the supercell level, providing sufficient degrees of freedom to attain full control over the five available diffraction channels at will. The interface, our design can successfully act as an anomalous reflector, retroreflector, and three-channel power splitter when excited from specific ports. The mission of the proposed metagrating can be switched by mere modulation of the DC biasing voltages. Compared with the similar studies, our design is the first all-graphene proposal with a simpler geometry while exposing a higher number of controllable channels for THz wave manipulation. Meanwhile, the proposed architecture could be straightforwardly extended to more complicated scenarios with larger number of diffraction modes through increasing the number of graphene ribbons in each period and seeking for proper set of chemical potentials. The proposed reconfigurable all-graphene metagrating may be a key in realizing highly-efficient THz meta-reflectors that could perform all possible transformations within a shared ultra-thin aperture.

**Disclosures.** The authors declare no conflicts of interest.

12. Popov, Vladislav, Fabrice Boust, and Shah Nawaz Burokur. "Controlling diffraction patterns with metagratings." Physical Review Applied 10, no. 1 (2018): 011002.
13. Asadchy, V. S., Ana Díaz-Rubio, S. N. Tcvetkova, D-H. Kwon, A. Elsakka, M. Albooyeh, and S. A. Tretyakov. "Flat engineered multichannel reflectors." Physical Review X 7, no. 3 (2017): 031046.
14. Ra'di, Younes, and Andrea Alù. "Reconfigurable metagratings." ACS Photonics 5, no. 5 (2018): 1779-1785.
15. Yin, Li-Zheng, Tie-Jun Huang, Feng-Yuan Han, Jiang-Yu Liu, and Pu-Kun Liu. "Terahertz multichannel metasurfaces with sparse unit cells." Optics letters 44, no. 7 (2019): 1556-1559.
16. Rouhi, Kasra, Hamid Rajabalipanah, and Ali Abdolali. "Multi-bit graphene-based bias-encoded metasurfaces for real-time terahertz wavefront shaping: From controllable orbital angular momentum generation toward arbitrary beam tailoring." Carbon 149 (2019): 125-138.
17. Momeni, Ali, Kasra Rouhi, Hamid Rajabalipanah, and Ali Abdolali. "An information theory-inspired strategy for design of re-programmable encrypted graphene-based coding metasurfaces at terahertz frequencies." Scientific reports 8, no. 1 (2018): 6200.
18. Rouhi, Kasra, Hamid Rajabalipanah, and Ali Abdolali. "Real-time and broadband terahertz wave scattering manipulation via polarization-insensitive conformal graphene-based coding metasurfaces." Annalen der Physik 530, no. 4 (2018): 1700310.
19. Su, Xiaopeng, Zeyong Wei, Chao Wu, Yang Long, and Hongqiang Li. "Negative reflection from metal/graphene plasmonic gratings." Optics letters 41, no. 2 (2016): 348-351.
20. Pozar, David M. *Microwave engineering*. John Wiley & Sons, 2009.
5